\documentclass{basi}
\usepackage {graphicx}
\usepackage {times}
\begin{document}
\title[Deep $J$-band imaging of QSO candidates with the HCT]{Deep $J$-band imaging of high redshift QSO candidates with the Himalayan Chandra Telescope} 
\author[Tomotsugu Goto \& Devendra Ojha ]%
       {Tomotsugu Goto\thanks{e-mail:tomo@ir.isas.jaxa.jp} \\ 
         Department of Infrared Astrophysics, Institute of Space and Astronautical Science (ISAS)\\
 Japan Aerospace Exploration Agency (JAXA), 3-1-1 Yoshinodai, Sagamihara, Kanagawa 229-8510, Japan \\
\newauthor Devendra Ojha \thanks{e-mail:ojha@tifr.res.in}\\
 Infrared Astronomy Group, Department of Astronomy and Astrophysics,\\
   Tata Institute  of  Fundamental Research, Homi Bhabha Road, Colaba, 
   Mumbai- 400 005, INDIA.
}
%\pubyear{2001}
%\volume{29}
%\setcounter{page}{17}
%\date{Received 2001 May 30; accepted 2001 June 07}
\maketitle
\label{firstpage}
\begin{abstract}
  High redshift QSOs (redshift$>$5.7) are highly important objects. If such QSOs may be found, their spectra will reveal the onset of reionization of the intergalactic medium (Gunn-Peterson trough),  and  provide precious insights into the re-ionization epoch in the very early universe. 
 Here we report our pilot attempt to follow-up high redshift QSOs with the  Himalayan Chandra Telescope. 
 Deep $J$-band imaging was performed on three high redshift QSO candidates color-selected from the SDSS, using the near-infrared imager.  Although none of the targets turned out to be likely high redshift QSOs, careful data reduction shows that the data reach the required depth, proving that the  Himalayan Chandra Telescope is a powerful tool to follow-up high redshift QSO candidates.
\end{abstract}

\section{Introduction}
\label{sec:intro}
High-redshift QSOs provide direct probes of the epoch
when the first generation of galaxies and QSOs formed.
The absorption spectra of these QSOs reveal the state of the intergalactic 
medium (IGM) close to the reionization epoch (Haiman et al. 1999; 
 % Miralda-Escud\'{e} et al. 2000, ApJ, 530, 1, 
 Madau et al. 2000; Cen et al. 2000).
The lack of a Gunn-Peterson trough %(Shklovsky 1964, Astron. Zh. 41, 408, Scheuer 1965, Nature, 207, 963, 
 (Gunn \& Peterson 1965) in the spectrum of the luminous QSO at $z=6.43$ (Fan et al. 2003) indicates that the universe was already highly ionized
at that redshift.
Assuming that the QSO is radiating at the Eddington luminosity, 
this object contains a central black hole of several billion solar masses (Fan et al. 2003).
The assembly of such massive objects in a timescale shorter than
1 gigayear yields constraints on models of the formation of massive black holes 
(e.g., Haiman et al. 2001).
The abundance and evolution of such QSOs can provide sensitive
tests for models of QSO and galaxy evolution. 
Therefore, high redshift QSOs are highly important objects in variety of scientific aspects.

\begin{figure*}
\begin{center}
\includegraphics[scale=0.45]{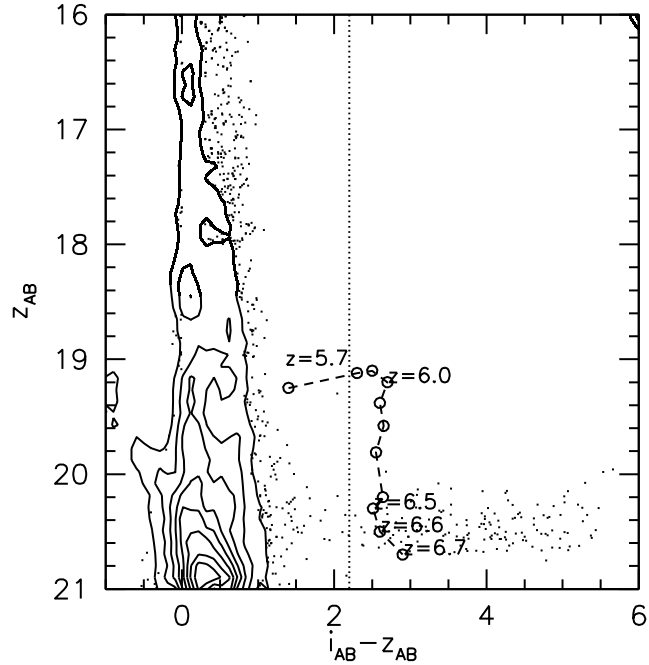}
\end{center}
\caption{The $i_{AB}-z_{AB}$ vs $z_{AB}$ color-color diagram for point
 sources in the SDSS. The median track of simulated  $i_{AB}-z_{AB}$
 color (taken from Fan et al. 2001) and $z_{AB}$ magnitude for QSOs with $M_{1450}=-27$ is shown in
 the dashed line as a function of redshift. 
 In the dense regions of stellar locus
 $i_{AB}-z_{AB}\sim 0.5$, dots are replaced with contours for
 clarity. 
 The color criteria of  $i_{AB}-z_{AB}=2.2$ is shown in the dotted line.
}\label{fig:z_iz}
\end{figure*}

\begin{figure*}
\begin{center}
\includegraphics[scale=0.3]{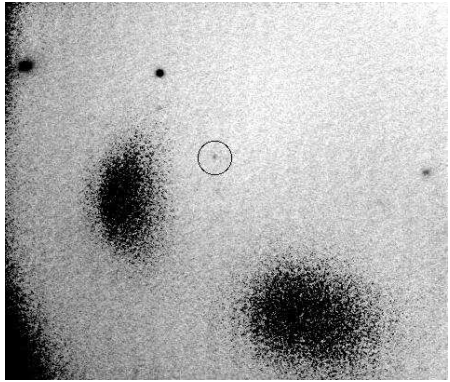}
\end{center}
\caption{The reduced image of the J0947 field. The QSO candidate was detected and circled in the image.
}\label{fig:image}
\end{figure*}

\section{Target Selection}
\label{sec:target_selection}
However, a search for such high redshift QSOs poses a big observational challenge. 
%However, a search for such high redshift QSOs challenges observer's toughness. 
 QSOs are rare objects. Due to the huge distance to them, only the bright QSOs can be observed from the currently available facilities. Therefore, the large volume must be searched to find such QSOs. 
We overcome this problem using the Sloan Digital Sky Survey (SDSS; Abazajian et al. 2004), which uses
a dedicated 2.5m telescope and a large format CCD camera %(Gunn et al. 1998 AJ, 116, 3040)
 %at the Apache Point Observatory in New Mexico
to obtain images in five optical broad bands %( Fukugita et al. 1996, AJ, 111, 1748)
 ($u$, $g$, $r$, $i$ and $z$ centred at 3551, 4686, 6166, 7480 and 8932 \AA, respectively; Fukugita et al. 1996) 
over 10,000 deg$^2$ of high Galactic latitude sky.
 This unprecedented large sky coverage %in $z$-band 
 provides us a unique
 opportunity to find a very rare class of objects such as high
 redshift QSOs, passive spiral galaxies (Goto et al. 2003a), and E+A galaxies (Goto el. 2003b; Goto 2004,2005). The inclusion of the reddest band, $z$, in principle enables the discovery of QSOs up to redshift$\sim 6.7$ from the SDSS data as a $z$-band only detection (Fan et al.2000). In this work, we have selected our targets from the fourth public data release of the SDSS (Abazajian et al. 2004) as follows.

 At redshift $>5.7$, the Ly$\alpha$ emission and the Lyman break of QSOs move into $z$-band, and due to the absorption from the neutral hydrogen, QSOs at redshift $>5.7$ should not have any flux in the $u,g,r,i$ bands in the SDSS. Therefore, we require targets not to  be detected in $u,g,r,i$ bands, and detected in $z$-bands ($z$-band only detection). 
 We require our targets to have the SDSS color of $i_{AB}-z_{AB}>2.2$. This color criterion is based on the median track of artificially redshifted known QSOs (see Fan et al. 2001; Chiu et al. 2005 for details). Objects in this color range is well-separated from that of main-sequence stars (Fan et al. 2001), and thus, this color-cut nicely removes the large contamination from stars.  We illustrate this color criteria in Figure \ref{fig:z_iz} as the dotted line. The median track of QSOs shown in the dashed line is well separated by the criteria from the stellar locus shown in the contour.   

 However, since high redshift QSOs are so rare, the remaining candidates still have large contamination from cosmic-rays and late-type stars, which have essentially the same optical color as high redshift QSOs.
 To remove these contaminants, $J$-band imaging with the 2-4m class telescopes is necessary.  
 Cosmicrays can be removed by obtaining multiple exposures.
 %Cosmic-rays can be completely removed by yet another imaging. 
 The $z-J$ color is a powerful separator of high-redshift QSOs ($z_{AB}-J_{AB}\leq 0.8$) from late-type stars ($z_{AB}-J_{AB}>0.8$). 

 In this paper, we report our pilot attempt to use the Himalayan Chandra telescope/the near infrared imager for this purpose. Once a promising high redshift QSO candidate with $z_{AB}-J_{AB}\leq 0.8$ is found, we plan to spectroscopically  follow-up with an 8m class telescope to confirm the discovery of a high redshift QSO (see Goto et al. 2006). 

\begin{figure*}
\begin{center}
\includegraphics[scale=0.3]{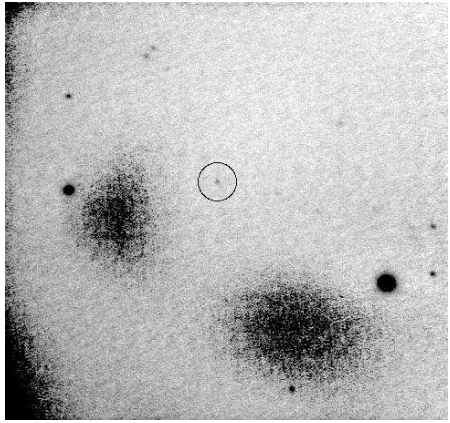}
\end{center}
\caption{The reduced image of the J1051 field. The QSO candidate was detected and circled in the image.
}\label{fig:J1051}
\end{figure*}

\section{Observation}
\label{sec:observation}

\begin{figure*}
\begin{center}
\includegraphics[scale=0.3]{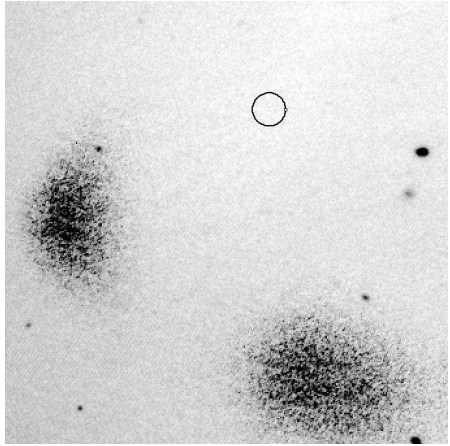}
\end{center}
\caption{The reduced image of the J1532 field.  The expected position of the QSO candidate is circled, but it was not detected. The 3 $\sigma$ detection limit is $J_{AB}$=21.2 mag.}\label{fig:J1532}
\end{figure*}

We have performed deep $J$-band imaging of high redshift QSO candidates with the near-infrared imager mounted on the 2m Himalayan Chandra Telescope on the nights of March 08 and April 18, 2006.  Although the telescope itself is at Digpa-ratsa Ri, Hanle, at 4500m of altitude in India, the observation was remotely performed from the Centre for Research and Education in Science and Technology (CREST) in Hosakote, via a dedicated satellite link.  Our goal here is to image as deep as $J_{AB} \sim$19.9 mag in order to distinguish high redshift QSO ($z_{AB}-J_{AB}\leq 0.8$) from late-type stars ($z_{AB}-J_{AB}>0.8$) for our candidates of $z_{AB}\sim$20.2 mag. 
Both nights were not photometric due to the presence of thin clouds. 
The coordinates and extinction-corrected optical magnitudes of targets are summarized in Table \ref{tab:targets}. Total exposure time was $\sim$1 hour for each target. The exposure was divided into multiple 90 sec exposures due to the high background in near infrared.  
The near infrared array is a 512 $\times$ 512 HgCdTe array of 18 $\mu$m pixel size. 
 For a larger field of view (3.6 $\times$ 3.6 arcmin$^2$), we have used the Camera B with 0.4 arcsec/pixel scale.
Unfortunately, the array has two potato shaped bad features in the left and bottom region (see Figs.\ref{fig:image}-\ref{fig:J1532}). The features are present at dark, sky and object frames, and cannot be taken out with the standard reduction procedures. The features are too large to be taken out with the dithering. Therefore,  to avoid these, we have placed our targets in the upper right region of the array. Due to the same reason, we have used small offsets of 5-10 arcsec in our 6-point dithering.

 The data were reduced using the IRAF package {\ttfamily upsqiid}\footnote{available at http://www.noao.edu/kpno/sqiid/upsqiidpkg.html}. We show the reduced images of the three targets in Figs.\ref{fig:image}-\ref{fig:J1532}. J0947 and J1051 are detected and marked with a black circle.
  Since both nights were not photometric, we used 2MASS stars in the field for photomeric calibration. 
 Typical 2MASS sources used for the photometric calibration have magnitude errors of $\sim$0.08 mag. This directly affects the zero-point accuracy of our HCT data. Therefore, we expect $\sim 0.1$mag of zero-point uncertainty in the $J$-band magnitude. 
 The first two targets, J0947 and J1051, are securely detected with greater than 10$\sigma$ significance. The $J$ magnitude in AB system is presented in Table \ref{tab:targets}. Unfortunately, these two objects are too bright to be a QSO in $J$ (c.f. $z_{AB}-J_{AB}\leq 0.8$ for QSOs). Therefore, perhaps, these are late-type stars. However, the 3$\sigma$ detection limit measured from the clean region of the data, is $J_{AB}$=19.9 and 20.1 mag. These observed detection limits successfully reaches the required depth to find high redshift QSOs from the SDSS. 
 
 On April 18th, we have observed 6 targets in addition. However, the weather was cloudy all through the night and no useful data were taken except for J1532. We show the reduced image of the J1532 field in Fig.\ref{fig:J1532} with the expected position of the QSO circled. However, the QSO candidate was not detected. The 3 $\sigma$ detection limit measured from the clean region of the image is $J_{AB}$=21.2 mag. This limits the $z_{AB}-J_{AB}$ color of the candidate to be $<-1.0$, which is too blue to be a typical high redshift QSO. Perhaps, this object is a cosmicray/spurious detection in the optical data. 

%\begin{table*}
% \centering
% \begin{minipage}{140mm}
%  \caption{List of Targets}\label{tab:targets}
%  \begin{tabular}{@{}crrcclclc@{}}
%  \hline
%%   Object     &            & \multicolumn{4}{c}{Flux density (Jy)%}
%   Object &  $i_{AB}^*$ & $z_{AB}^*$ & $J_{Vega}$  & 3$\sigma$ detection limit & Exposure time (min) & Observing dates   & \\
%%        &  &  &  &  &  & group & (d) & curve \\
%%        &  &  &  &  &  &       &     & type  \\
% \hline
%J094744.71+414622.2 &  22.60 & 20.15 & 17.9 ($>$10$\sigma$) & 19.2 &54 (90sec$\times$36) & 08, March, 2006   &  &   \\
%J105124.31+541401.5 &  22.91 & 20.11 & 18.0 ($>$10$\sigma$) & 19.0 &54 (90sec$\times$36) & 08, March, 2006   &  &   \\
%%J1527 & 5.96 & 22.54 & 19.84 & 18.9 & 63 (90sec$\times$42) & 18, April, 2006   &  &  & \\星が一個写っているのみ。雲がかかっていたのか？　
%J153242.19+234418.1 &  24.24 & 20.20 & not detected & 20.3 & 63 (90sec$\times$42) & 18, April, 2006   &  &   \\
%\hline
%\end{tabular}
%\end{minipage}
%\end{table*}

%J in AB. add 0.908 to Vega.

\begin{table*}
 \centering
 \begin{minipage}{140mm}
  \caption{List of Targets}\label{tab:targets}
  \begin{tabular}{@{}crrcclclc@{}}
  \hline
%   Object     &            & \multicolumn{4}{c}{Flux density (Jy)%}
   Object &  $i_{AB}^*$ & $z_{AB}^*$ & $J_{AB}$  & 3$\sigma$ detection limit & Exposure time (min) & Observing dates   & \\
%        &  &  &  &  &  & group & (d) & curve \\
%        &  &  &  &  &  &       &     & type  \\
 \hline
J094744.71+414622.2 &  22.60 & 20.15 & 18.8 ($>$10$\sigma$) & 20.1 &54 (90sec$\times$36) & 08, March, 2006   &  &   \\
J105124.31+541401.5 &  22.91 & 20.11 & 18.9 ($>$10$\sigma$) & 19.9 &54 (90sec$\times$36) & 08, March, 2006   &  &   \\
%J1527 & 5.96 & 22.54 & 19.84 & 18.9 & 63 (90sec$\times$42) & 18, April, 2006   &  &  & \\星が一個写っているのみ。雲がかかっていたのか？　
J153242.19+234418.1 &  24.24 & 20.20 & not detected & 21.2 & 63 (90sec$\times$42) & 18, April, 2006   &  &   \\
\hline
\end{tabular}
\end{minipage}
\end{table*}

\section{Summary}

As a pilot study, we have imaged three high redshift QSO candidates using the Himalayan Chandra telescope in $J$-band. Two of the targets are likely to be late-type stars, and the other is perhaps spurious detection in the optical data. However, all of the three fields reached the required depth of $J_{AB}>$19.9 mag to find a high redshift QSOs from the SDSS. Our results show that the  Himalayan Chandra telescope is a useful tool to follow-up high redshift QSO candidates. % if the weather permits.

%\section{Bibliography and the rest}
%All other commands, including bibliography formatting and citing, work
%as in standard article style.  No special command has been introduced 
%for formatting the list of references, they should appear, as usual in
%the\\ 
%\verb+\thebibliography+\\ 
%environment, each entry as an individual \\
%\verb+\bibitem+.\\  
%However, a few changes have been made in the output
%format of the references to suit the BASI style.  
%For example, it is set in a relatively smaller font than the text, and
%the serial number of the references are omitted.  However, for compatibility
%reasons the \verb+\thebibliography+ command must continue to carry a
%numeric argument, which is traditionally used to choose the width of
%the label field.
%

\section*{Acknowledgements}
 We thank the anonymous refereee for many insightful comments, which
 improved the paper significantly. 
T.G. thanks the Foundation for Promotion of Astronomy for the financial support for the observing travel.

%Writing the file \verb+basi.sty+ has been prompted by the numerous
%requests I received for something of this form during the years I 
%spent as the secretary of the Astronomical Society of India.  However
%it finally saw the light of the day when it became necessary to process
%some of my own papers rather quickly, and I found that electronic
%typesetting using this style file would be a way to substantially 
%reduce the production time.  I am grateful to the large community of
%authors for demanding a \LaTeX\ style file, and to the Editor Harish
%C. Bhatt for encouraging me to make this available.  Thanks are due
%to Sandra Rajiva for checking through the outputs carefully to
%ensure that the formatting style is correct in detail.

%Some fraction of the code of \verb+basi.sty+ is common with the
%style file \verb+jaa.sty+ that I had prepared in 1996 for the 
%Journal of Astrophysics of Astronomy, published by the Indian Academy
%of Sciences.  A specific piece of code, for author/address formatting, 
%has been adapted from mn.sty (copyright Cambridge University Press),
%with minor modifications.  
\label{lastpage}
\end{document}